\documentclass[twocolumn,showpacs,preprintnumbers]{revtex4}
\usepackage{amssymb}
\usepackage{amsfonts}
\usepackage{amsmath}
\usepackage{graphicx}
\usepackage{dcolumn}
\usepackage{bm}

\input{tcilatex}

\begin{document}

\title{ Beam Splitter Entangler for Light Fields }
\author{Xun-Li Feng and Zhi-Zhan Xu}
\affiliation{Laboratory for High Intensity Optics, Shanghai Institute of Optics and Fine
Mechanics, \\
Chinese Academy of Sciences, Shanghai, 201800, People's Republic of China}

\begin{abstract}
We propose an experimentally feasible scheme to generate various types of
entangled states of light fields by using beam splitters and single-photon
detectors. Two light fields are incident on two beam splitters and are split
into strong and weak output modes respectively. A conditional joint
measurement on both weak output modes may result in an entanglement between
the two strong output modes. The conditions for the maximal entanglement are
discussed based on the concurrence. Several specific examples are also
examined.
\end{abstract}

\maketitle

%\preprint{LE9647.Feng}

Quantum entanglement has been identified as a basic resource in achieving
tasks of quantum communication and quantum computation \cite{1}. Photons are
considered to be the best quantum information carriers over long distances,
and those in entangled states have been used to experimentally demonstrate
quantum teleportation \cite{2}, quantum dense coding \cite{3}, quantum
cryptography \cite{4}, and the generation of GHZ-states of three or four
photons \cite{5}.

In comparison with other candidates for engineering quantum entanglement,
light fields possess more abundant capacity to create various types of
entangled states including the discrete, the continuous variable, and the
combination of the both. Almost all of the reported experiments adopted the
parametric down-conversion process as a standard source for the entangled
photon pairs as well as two-mode squeezed states. Recently it was found that
linear optical elements, such as beam-splitters and polarization
beam-splitters, can also be used to generate entangled light fields \cite%
{6,7,8,9,10,11}. In Ref. \cite{6}, an efficient quantum computation with
linear optics is devised, it is undoubtedly an entangler for light fields.
On the other hand, a single beam splitter can also act as an entangler for
light fields if the input modes are in appropriate nonclassical states \cite%
{10,11}. However, the types of the resultant entangled states from the
existent schemes are still very limited and cannot satisfy the requirements
for quantum information processing and for other applications. Thus how to
generate various types of entangled states of light fields is still of great
significance.

Very recently several schemes have been proposed to entangle distant atoms 
\cite{12} and atomic ensembles \cite{13} by means of photon interference. It
is natural to ask whether the idea of photon interference can be extended to
the generation of\ the entangled states of light fields. This letter will
give a positive answer.

In this letter we present an experimentally feasible scheme for the
generation of entangled states of light fields by using beam splitters and
the single-photon detectors. The setup we are considering here consists of
three lossless beam splitters, BS$_{1}$, BS$_{2}$ and BS$_{3}$, and two
single-photon detectors D$_{1}$ and D$_{2}$, as shown in Fig. 1. For
simplicity, we assume that BS$_{1}$ and BS$_{2}$ are of the same amplitude
reflection and transmission coefficients $R$ and $T$ with a relation $%
|R|^{2}+|T|^{2}=1,$ and that the distance from BS$_{1}$ to BS$_{3}$ is the
same as that from BS$_{2}$ to BS$_{3}$. As an optical element, a beam
splitter generally has two input ports and two output ports. In our setup
only one input port has a non-zero input field for both BS$_{1}$ and BS$_{2}$%
, the other input port is always left in the vacuum state. In this case the
non-zero input field will be split into two output fields. As has been
proved, the two output fields may be entangled when the input field is in an
appropriate nonclassical state \cite{10,11}. The entanglement quantity and
the properties of the output fields depend to a large extent on the input
fields as well as on the amplitude reflection and transmission coefficients $%
R$ and $T$ of the beam splitter. If $R$ and $T$ of the beam splitter BS$_{1}$
(and BS$_{2}$) are largely different, the non-zero input field will be split
into two output fields in which one is very strong, and the other is very
weak. In fact, we can design BS$_{1}$ and BS$_{2}$ in such a way that the
weak output field possesses maximally one photon, that is, the weak output
mode is in\ the vacuum state $|0\rangle $\ or in the one photon number state 
$|1\rangle .$ Correspondingly the strong output mode will be in a state that
keeps the photon number of the input field unchanged or in a state that
annihilates one photon from the input field since the lossless beam splitter
conserves the photon number of the input fields. In this way we have
prepared a specific entangled state of the weak and the strong output
fields. Subsequently we let the two weak output fields from BS$_{1}$ and BS$%
_{2}$ be combined at BS$_{3}$, a 50\%:50\% beam splitter, and then detected
by single-photon detectors D$_{1}$ and D$_{2}$. If only one photon is
registered by D$_{1}$ or D$_{2}$, we successfully generate an entangled
state of the two strong output fields of BS$_{1}$ and BS$_{2}$. Otherwise,
we fail to generate the desired entangled state and should repeat the
process again until one photon is registered.

Our scheme is based on a simple optical element, the beam splitter. The
perfection of the beam splitter is very high, the requirements for BS$_{1}$
and BS$_{2}$ can thus be satisfied with the required precision easily,
therefore our scheme is experimentally feasible. In particular, as will be
shown below, our scheme can be used to generate various types of entangled
states of light fields.

In order to illustrate our method explicitly, we first examine the effect of
a beam splitter on its output fields. Let us denote by $\hat{a}_{j},\hat{b}%
_{j}$ the input mode amplitudes shown in Fig. 1 and by $\hat{c}_{j},\hat{d}%
_{j}$ the output mode amplitudes, where the subscript $j$ stands for BS$_{j}$%
($j=1,2$). Suppose the initial quantum state of the input fields for BS$_{j}$
is a product state, $|0,\psi _{j}\rangle _{in}=|0\rangle _{a_{j}}\otimes
|\psi _{i}\rangle _{b_{j}}$, in which the mode $\hat{a}_{j}$ is supposed to
be always in the vacuum state $|0\rangle _{a_{j}}$ and the mode\ $\hat{b}%
_{j} $ in a superposition of the photon number states which can be expressed
as

\begin{equation}
|\psi _{j}\rangle =\sum_{n=0}^{\infty }f_{n}^{(j)}|n\rangle ,  \tag{1}
\end{equation}%
with the normalized condition $\sum_{n=0}^{\infty }|f_{n}^{(j)}|^{2}=1$. The
output fields are then of the following form \cite{14}, 
\begin{eqnarray}
|\Psi _{j}\rangle _{out} &=&\hat{B}_{j}|0,\psi _{j}\rangle _{in}  \notag \\
&=&\sum_{n=0}^{\infty }f_{n}^{(j)}\hat{B}_{j}|0,n\rangle _{in},  \TCItag{2}
\end{eqnarray}%
where the beam splitter operator is 
\begin{equation}
\hat{B}_{j}=\exp \left[ {\frac{\theta }{2}}(\hat{a}_{j}^{\dag }\hat{b}%
_{j}e^{i\phi }-\hat{a}_{j}\hat{b}_{j}^{\dag }e^{-i\phi })\right] ,  \tag{3}
\end{equation}%
with the amplitude reflection and transmission coefficients $T=\cos {\frac{%
\theta }{2}},~~R=\sin {\frac{\theta }{2}}${.} Here $\phi $ denotes the phase
difference between the reflected and transmitted fields.

The terms $\hat{B}_{j}|0,n\rangle _{in}$ in Eq. (2) can be evaluated
according to Eq. (3)$,$ 
\begin{equation}
\hat{B}|0,n\rangle _{in}=\sum_{k=0}^{n}c_{k}^{n}|k,n-k\rangle _{out}, 
\tag{4}
\end{equation}%
where $|k,n-k\rangle _{out}=|k\rangle _{c}\otimes |n-k\rangle _{d}$, and $%
c_{k}^{n}=\binom{n}{k}^{1/2}\exp \left( ik\phi \right) R^{k}T^{n-k}.$

In the following we assume that the amplitude reflection coefficient $R$ for
BS$_{j}$ ($j=1,2$) is so small ($R\ll T$) that the terms containing $R^{k}$
in Eq. (4) can be neglected when $k\geqslant 2$, thus Eq. (4) is simplified
as

\begin{eqnarray}
&&\hat{B}_{j}|0,n\rangle _{in}  \notag \\
&\approx &T^{n}|0,n\rangle _{out}+\exp \left( i\phi \right) RT^{n-1}\sqrt{n}%
|1,n-1\rangle _{out}  \notag \\
&=&T^{n}|0,n\rangle _{out}+\exp \left( i\phi \right) RT^{n-1}\hat{d}%
_{j}|1,n\rangle _{out}.  \TCItag{5}
\end{eqnarray}%
Substituting Eq. (5) into Eq. (2) we get

\begin{equation}
|\Psi _{j}\rangle _{out}=|0\rangle _{c_{j}}\otimes |u_{j}\rangle
_{d_{j}}+\exp \left( i\phi \right) R|1\rangle _{c_{j}}\otimes |v_{j}\rangle
_{d_{j}},  \tag{6}
\end{equation}%
where $|u_{j}\rangle $ and $|v_{j}\rangle $ take the form

\begin{eqnarray}
|u_{j}\rangle &=&\sum_{n=0}^{\infty }f_{n}^{(j)}T^{n}|n\rangle ,  \TCItag{7a}
\\
|v_{j}\rangle &=&\hat{d}_{j}\sum_{n=0}^{\infty
}f_{n+1}^{(j)}T^{n}|n+1\rangle .  \TCItag{7b}
\end{eqnarray}%
The state (6) is a specific entangled state in which the mode $\hat{c}_{j}$ (%
\textit{i.e.} the weak mode) possesses maximally one photon.

Now we show how to generate the entangled states of two strong output modes $%
\hat{d}_{1}$ and $\hat{d}_{2}$ by manipulating the two weak modes $\hat{c}%
_{1}$ and $\hat{c}_{2}$. As shown in Fig. 1, we suppose the two input fields
are, respectively, incident on BS$_{1}$ and BS$_{2}$ simultaneously.
Afterwards the two weak output modes $\hat{c}_{1}$ and $\hat{c}_{2}$ are
combined at BS$_{3}$ with the output mode amplitudes $\hat{A}_{1}=\left[ i%
\hat{c}_{1}+e^{i\gamma }\hat{c}_{2}\right] /\sqrt{2}$,$~\hat{A}_{2}=\left[ 
\hat{c}_{1}+ie^{i\gamma }\hat{c}_{2}\right] /\sqrt{2}$. Here the factor $i$
denotes a $\pi /2$ phase shift between the reflected and the transmitted
modes, and $\gamma $ denotes the phase shift induced by a wave plate (WP)
placed in the path of mode $\hat{c}_{2}$. The detection of a single photon
by D$_{1}$ or D$_{2}$ is accompanied by the wave function collapse $|\Psi
_{1}\rangle _{out}\otimes |\Psi _{2}\rangle _{out}\rightarrow \hat{A}%
_{1,2}(|\Psi _{1}\rangle _{out}\otimes |\Psi _{2}\rangle _{out})$.
Neglecting the high-order terms $o(R^{2}),$ we find the final state of the
two strong modes, conditional on a click of either D$_{1}$ or D$_{2}$, takes
the following form, 
\begin{equation}
|\Phi \rangle _{1,2}=|v_{1}\rangle |u_{2}\rangle \mp i\exp \left( i\gamma
\right) |u_{1}\rangle |v_{2}\rangle .  \tag{8}
\end{equation}%
Here and in what follows the subscripts $d_{j}$ are omitted to simplify the
notation. Note that the state (8) is not normalized. The probability for
successfully generating the above state is proportional to $R^{2}$. The
phase factor in state (8) can be controlled through the WP.

In principle, the scheme we proposed here is closely related to the quantum
entanglement swapping \cite{15} in which the particles that have never
interacted directly are entangled, and can be regarded as a specific version
of entanglement concentration protocol by using entanglement swapping \cite%
{16}. The entanglement between the weak and the strong output fields for
both BS$_{1}$ and BS$_{2}$ is a necessary condition for the successful
generation of the desired entangled state (8). Accordingly the requirement
that both input fields of BS$_{1}$ and BS$_{2}$ are in nonclassical states
should be satisfied \cite{10,11}.

Now let us examine the entanglement properties of the resultant state (8).
To this end, we first transform it to the normalized basis,

\begin{equation}
|\Phi \rangle _{1,2}=\nu _{1}\mu _{2}|V_{1}\rangle |U_{2}\rangle \mp i\exp
\left( i\gamma \right) \mu _{1}\nu _{2}|U_{1}\rangle |V_{2}\rangle ,  \tag{9}
\end{equation}%
where $|U_{j}\rangle =|u_{j}\rangle /\mu _{j}$, $|V_{j}\rangle
=|v_{j}\rangle /\nu _{j}$ are normalized states for system $j$ with the
normalized constants

\begin{eqnarray}
\mu _{j} &=&\left( \sum_{n=0}^{\infty }|T^{n}f_{n}^{(j)}|^{2}\right) ^{\frac{%
1}{2}},  \TCItag{10a} \\
\nu _{j} &=&\left( \sum_{n=0}^{\infty }\left( n+1\right)
|T^{n}f_{n+1}^{(j)}|^{2}\right) ^{\frac{1}{2}}.  \TCItag{10b}
\end{eqnarray}%
Considering $|U_{j}\rangle $ and $|V_{j}\rangle $ may be nonorthogonal, the
state (9) is actually a general bipartite entangled state. The concurrence 
\cite{17} has been proved to be a convenient entanglement measure for such
states and has been evaluated by Wang \cite{18}. We find the concurrence of
the entangled state (9) takes the form 
\begin{widetext}
\begin{equation}
C_{1,2}=\frac{2\mu _{1}\nu _{1}\mu _{2}\nu _{2}\sqrt{\left( 1-|\langle
V_{1}|U_{1}\rangle |^{2}\right) \left( 1-|\langle V_{2}|U_{2}\rangle
|^{2}\right) }}{\mu _{1}^{2}\nu _{2}^{2}+\nu _{1}^{2}\mu _{2}^{2}+\mu
_{1}\nu _{1}\mu _{2}\nu _{2}\left( \pm i\text{e}^{-i\gamma }\langle
U_{1}|V_{1}\rangle \langle V_{2}|U_{2}\rangle \mp i\text{e}^{i\gamma
}\langle V_{1}|U_{1}\rangle \langle U_{2}|V_{2}\rangle \right) }.  \tag{11}
\end{equation}%
\end{widetext}The concurrence ranges from 0 to 1 with the value 1
corresponding to a maximally entangled state (MES). In a special case where
the input field of BS$_{1}$ is the same as that of BS$_{2}$,\textit{\ i.e., }%
$f_{n}^{(1)}=f_{n}^{(2)}$, we have $|u_{1}\rangle =|u_{2}\rangle \equiv
|u\rangle $, $|v_{1}\rangle =|v_{2}\rangle \equiv |v\rangle $, $\mu _{1}=\mu
_{2}$, $\nu _{1}=\nu _{2}$, and therefore $|U_{1}\rangle =|U_{2}\rangle
\equiv |U\rangle $, $|V_{1}\rangle =|V_{2}\rangle \equiv |V\rangle $, the
concurrence is thus simplified as 
\begin{equation}
C_{1,2}=\frac{1-|\langle V|U\rangle |^{2}}{1\pm \sin \gamma |\langle
V|U\rangle |^{2}}.  \tag{12}
\end{equation}%
From Eq. (12) one can readily deduce the following conclussion: (i) If $%
|U\rangle \ $and $|V\rangle $ are orthogonal, i.e., $\langle V|U\rangle
=\langle v|u\rangle =0$, we always have $C_{1,2}=1$, and the resultant
entangled state (8) is an MES. (ii)\ If, however, $|U\rangle \ $and $%
|V\rangle $ are nonorthogonal, the condition that state (8) is an MES is $%
\gamma =3\pi /2$ when the photon is detected by D$_{1}$ or $\gamma =\pi /2$
when the photon is detected by D$_{2}$.

Finally we give several examples to demonstrate the power of our scheme as
an entangler for the generation of various types of entangled states of
light fields.

\textit{Example 1, photon number state inputs.} When the input field of BS$%
_{1}$ and that of BS$_{2}$ are in the Fock states $|n\rangle $ and $%
|m\rangle $ respectively, the resultant entangled state is 
\begin{eqnarray}
|\Phi \rangle _{1,2} &=&\frac{1}{\sqrt{n+m}}\left[ \sqrt{n}|n-1\rangle
|m\rangle \right.  \notag \\
&&\left. \mp i\exp \left( i\gamma \right) \sqrt{m}|n\rangle |m-1\rangle %
\right] .  \TCItag{13}
\end{eqnarray}%
If $n=m$, the above state becomes an MES

\begin{equation}
|\Phi \rangle _{1,2}=\frac{1}{\sqrt{2}}\left[ |n-1\rangle |n\rangle \mp
i\exp \left( i\gamma \right) |n\rangle |n-1\rangle \right] .  \tag{14}
\end{equation}

\textit{Example 2},\textit{\ even (or odd) coherent state inputs.} Even and
odd coherent states are superposition of coherent states, \textit{i.e. }Shr%
\"{o}dinger cat states, and take the following form

\begin{eqnarray}
|\alpha \rangle ^{e} &=&N_{e}(|\alpha \rangle +|-\alpha \rangle )  \notag \\
&=&\left( \cosh |\alpha |^{2}\right) ^{-\frac{1}{2}}\sum_{n=0}^{\infty }%
\frac{\alpha ^{2n}}{\sqrt{(2n)!}}|2n\rangle ,  \TCItag{15a} \\
|\alpha \rangle ^{o} &=&N_{o}(|\alpha \rangle -|-\alpha \rangle )  \notag \\
&=&\left( \sinh |\alpha |^{2}\right) ^{-\frac{1}{2}}\sum_{n=0}^{\infty }%
\frac{\alpha ^{2n+1}}{\sqrt{(2n+1)!}}|2n+1\rangle ,  \TCItag{15b}
\end{eqnarray}%
where $|\alpha \rangle \ $and $|-\alpha \rangle $ are coherent states, $%
N_{e} $ and $N_{o}$ are normalized coefficients. If both input field of BS$%
_{1}$ and that of BS$_{2}$ are in the same even coherent states, $|\alpha
\rangle ^{e}$, the resultant entangled state is 
\begin{equation}
|\Phi \rangle _{1,2}=\frac{1}{\sqrt{2}}\left[ |T\alpha \rangle ^{o}|T\alpha
\rangle ^{e}\mp i\exp \left( i\gamma \right) |T\alpha \rangle ^{e}|T\alpha
\rangle ^{o}\right] .  \tag{16}
\end{equation}%
It is well known that even and odd coherent states are orthogonal, the
resultant state (16) is thus an MES. A similar result can be worked out if
both input field of BS$_{1}$ and that of BS$_{2}$ are in the same odd
coherent states, $|\alpha \rangle ^{o}$.

\textit{Example 3, squeezed vacuum state inputs. }A single mode squeezed
vacuum state is a superposition of even photon number states,

\begin{equation}
|SV(r)\rangle =\left( \cosh r\right) ^{-1/2}\sum_{n=0}^{\infty }\frac{\left(
-\Gamma \right) ^{n}\sqrt{(2n)!}}{n!2^{n}}|2n\rangle ,  \tag{17}
\end{equation}%
where $\Gamma =\exp (i\Theta )\tanh r,$ $r$ is the squeeze parameter. If
both input field of BS$_{1}$ and that of BS$_{2}$ are in the same squeezed
vacuum states, $|SV(r)\rangle $, the resultant entangled state is

\begin{eqnarray}
|\Phi \rangle _{1,2} &=&\hat{d}_{1}|SV(\bar{r})\rangle _{d_{1}}\otimes |SV(%
\bar{r})\rangle _{d_{2}}  \notag \\
&&\mp i\exp \left( i\gamma \right) |SV(\bar{r})\rangle _{d_{1}}\otimes \hat{d%
}_{2}|SV(\bar{r})\rangle _{d_{2}},  \TCItag{18}
\end{eqnarray}%
where $|SV(\bar{r})\rangle $ is a squeezed vacuum state with the squeezing
parameter $\bar{r}$ satisfying a relation $\tanh \bar{r}=T^{2}\tanh r.$
Because $\hat{d}_{j}|SV(\bar{r})\rangle _{d_{j}},$ which is a superposition
of the odd photon number states, and $|SV(\bar{r})\rangle _{d_{j}}$ are
orthogonal, the resultant state (18) is an MES.

In the above examples the resultant entangled states are either a discrete
entanglement (example 1) or continuous variable entanglement (examples 2 and
3). In practice, our scheme can be used to generate a combination
entanglement of the discrete and the continuous variable, see the following
example.

\textit{Example 4, Fock state and even coherent state inputs. }If one input
field, say, the input of BS$_{1}$, is in a Fock state $|n\rangle ,$ the
other input field, the input of BS$_{2}$, is in an even coherent states, $%
|\alpha \rangle ^{e}$, we may obtain the following entangled state 
\begin{eqnarray}
|\Phi \rangle _{1,2} &=&\left[ n\cosh \left( |T\alpha |^{2}\right) \right] ^{%
\frac{1}{2}}|n-1\rangle |T\alpha \rangle ^{e}\mp i\exp \left( i\gamma \right)
\notag \\
&&\times T\alpha \left[ \sinh \left( |T\alpha |^{2}\right) \right] ^{\frac{1%
}{2}}|n\rangle |T\alpha \rangle ^{o}.  \TCItag{19}
\end{eqnarray}%
The concurrence of the entangled state (19) is

\begin{equation}
C=\frac{2|T\alpha |\left[ n\sinh \left( |T\alpha |^{2}\right) \cosh \left(
|T\alpha |^{2}\right) \right] ^{1/2}}{|T\alpha |^{2}\sinh \left( |T\alpha
|^{2}\right) +n\cosh \left( |T\alpha |^{2}\right) }.  \tag{20}
\end{equation}%
If $|T\alpha |^{2}\tanh \left( |T\alpha |^{2}\right) =n,$ we have $C=1,$ and
the entangled state (19) is an MES.

In summary, we have presented an experimentally feasible scheme to generate
various types of entangled states of light fields by using beam splitters
and single-photon detectors. Our scheme is experimentally feasible because
the basic elements in our scheme are accessible to experimental
investigation with current technology. It has been shown that various types
of light fields, such as the discrete, the continuous variable light fields
and the combination of the both, can be entangled using our scheme.

{\Large Acknowledgments}

This work was supported by the National Key Basic Research Special
Foundation of China (Grant No. G1999075200) and the Natural Science
Foundation of China (Grant No. 60178014).

\begin{figure}[tbp]
\scalebox{0.4}{\includegraphics{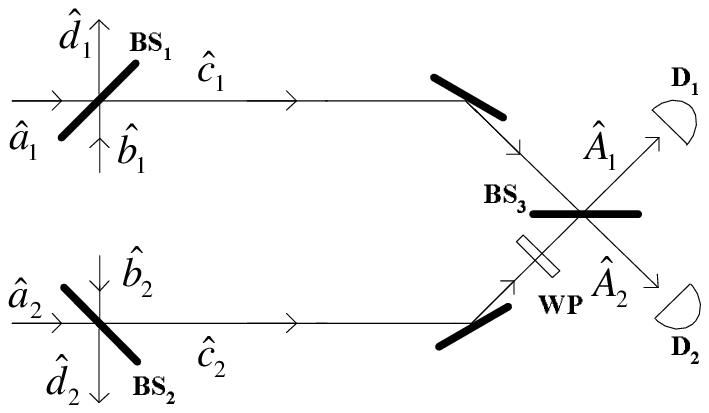}}
\caption{Experimental setup. Two input light fields in modes $\hat{b}_{1}$
and $\hat{b}_{2}$ are incident on BS$_{1}$ and BS$_{2},$ and split into weak
outputs in modes $\hat{c}_{1}$ and $\hat{c}_{2}$ and strong outputs in modes 
$\hat{d}_{1}$ and $\hat{d}_{2}$. Photons in weak outputs are then mixed on BS%
$_{3}$ and subsequently detected by single-photon detectors D$_{1}$ and D$%
_{2}$.}
\label{fig1}
\end{figure}

\end{document}